\documentclass{aa} 
\usepackage{natbib}
\usepackage[version=3]{mhchem}
\usepackage[figuresright]{rotating}

\bibpunct{(}{)}{;}{a}{}{,}

\usepackage{graphicx}
\usepackage{txfonts}

\begin{document} 

\title{Formation of COMs through CO hydrogenation on interstellar grains}\titlerunning{Formation of COMs on interstellar grains}
\author{M.A.J.~Simons\inst{1}
        \and
        T.~Lamberts\inst{2}
        \and
        H.M.~Cuppen\inst{1}
          }
   \institute{Radboud University Nijmegen, Institute for Molecules and Materials, Heyendaalseweg 135, 6525 AJ Nijmegen, The Netherlands\\                                                    
              \email{hcuppen@science.ru.nl}                                                                                                                                                                            
             \and
             Leiden Institute of Chemistry, Gorlaeus Laboratories, Leiden University, P.O. Box 9502, 2300 RA Leiden, The Netherlands                                                                                                                                                                          
             }                                                                                                                                                                                                         
   \date{Received \ldots; accepted \ldots}                                                                                                                                                                             
                                                                                                                                                                                                                      
\abstract                                                                                                                                                                                                              
{
Glycoaldehyde, ethylene glycol, and methyl formate are complex organic molecules that have been observed in dark molecular clouds. Because there is no efficient gas-phase route to produce these species, it is expected that a low-temperature surface route existst that does not require energetic processing. CO hydrogenation experiments at low temperatures showed that this is indeed the case. Glyoxal can form through recombination of two HCO radicals and is then further hydrogenated.}{
Here we aim to constrain the methyl formate, glycolaldehyde, and ethylene glycol formation on the surface of interstellar dust grains through this cold and dark formation route. We also  probe the dependence of the grain mantle composition on the initial gas-phase composition and the dust temperature. 
}{
A full CO hydrogenation reaction network was built based on quantum chemical calculations for  the rate constants and branching ratios. This network was used in combination with a microscopic kinetic Monte Carlo simulation to simulate ice chemistry, taking into account all positional information. After benchmarking the model against \ce{CO}-hydrogenation experiments, simulations under molecular cloud conditions were performed.
}{
Glycoaldehyde, ethylene glycol, and methyl formate are formed in all interstellar conditions we studied, even at temperatures as low as 8~K. This is because the \ce{HCO + HCO} reaction can occur when HCO radicals are formed close to each other and do not require to diffuse. Relatively low abundances of methyl formate are formed. The final COM abundances depend more on the H-to-CO ratio and less on temperature. Only above 16~K, where CO build-up is less efficient, does temperature start to play a role. Molecular hydrogen is predominantly formed through abstraction reactions on the surface. The most  important reaction leading to methanol is \ce{H2CO + CH3O -> HCO + CH3OH}. Our simulations are in agreement with observed COM ratios for mantles that have been formed at low temperatures. 
}{
}                                                                                                                                                                                                                      
                                                                                                                                                                                                                       
\keywords{  Astrochemistry --                                                                                                                                                                                          
            ISM: clouds --                                                                                                                                                                                             
            ISM: molecules --                                                                                                                                                                                          
            Methods: numerical --                                                                                                                                                                                      
               }                                                                                                                                                                                                       
                                                                                                                                                                                                                       
\maketitle

\section{Introduction}
About 200 molecules have been identified in interstellar space (see {\em https://cdms.astro.uni-koeln.de/classic/molecules}). Approximately 50 of them are classified as complex organic molecules (COMs), that is, they are\emph{} comprised of six or more atoms. COMs are generally observed in the gas phase, although it is widely accepted that saturated COMs are formed primarily through reactions on the surface of icy dust grains. This is where gas-phase species accrete, meet, and react to form saturated molecules. In the present paper we focus on the formation of glycolaldehyde and ethylene glycol. Glycoaldehyde and ethylene glycol have been detected towards the class 0 protostellar binary IRAS 1629-32422 \citep{Jorgensen:2012} and towards Galactic center source Sagittarius B2 \citep{Hollis:2000,Hollis:2002}.
These COMs are potential prebiotic molecules because they are precursor molecules for sugars, and they serve as an excellent test case for more complex grain surface chemistry networks because their formation involves the creation of a \ce{C-C} bond. 

The precise mechanism of COM formation is unknown, but is subject of many studies. Several different mechanisms have been suggested over the past decades. \citet{Charnley:1997}, \citet{Charnley:2005}, and \citet{Charnley:2009} applied a simple atom addition mechanism where species such as methanol can react with atomic carbon and are subsequently hydrogenated. Another mechanism was proposed by \citet{Garrod:2006a} and is more efficient at elevated temperatures. Cosmic-ray-induced photons can dissociate \ce{H2CO} and \ce{CH3OH}, creating functional-group radicals such as \ce{CH3} and \ce{CH3O} on or within the ice mantle. At low temperatures, these radicals are hydrogenated again, but as the temperature in the core increases to above $20-30$~K, the residence time of H atoms on the surface decreases substantially while at the same time radicals become mobile, thereby allowing radical-radical association reactions to become competitive with hydrogenation of radicals. Later work indeed showed that the formation of COMs, including glycolaldehyde and methyl formate, is possible at temperatures above 20~K \citep{Garrod:2008b, Vasyunin:2013}. Additionally, \citet{Laas:2011} showed that the branching ratio of radical formation affects the abundance of the recombined species as well.

The detection of acetaldehyde \ce{CH3CHO}, dimethyl ether \ce{CH3OCH3}, methyl formate \ce{CH3OCHO}, and ketene \ce{CH2CO} in the prestellar core L1689B suggests that COM synthesis has already started at the prestellar stage and a viable grain surface reaction route exists that is possible at low temperatures and does not require extensive UV irradiation \citep{Bacmann:2012}. The scenario described above requires both and therefore cannot be the only mechanism leading to the formation of COMs. Gas-phase routes have been proposed \citep{Skouteris:2018} as well as grain surface formation through cosmic-ray irradiation \citep{Shingledecker:2018,Shingledecker:2019}. 
\citet{Woods:2013} proposed a mechanism for the formation of COMs, where the essential carbon-backbone elongation occurs by recombination of two formyl radicals (HCO). Their astrophysical model matches the observed estimates in the hot molecular core G31.41+0.31 and low-mass binary protostar IRAS 1629-32422. 
Recently, \citet{Fedoseev:2015} confirmed this picture by showing that CO-hydrogenation experiments yield glycoaldehyde and ethylene glycol at 13~K. The authors argued that the formation of the \ce{C-C} bond must have occurred through a recombination of two HCO radicals although the product glyoxal (\ce{HC(O)CHO}) was not detected because other proposed mechanisms should have an activation barrier and are thus not feasible at low temperatures.
This is an important experimental finding because surface hydrogenation reactions have so far mainly been shown to be effective in the formation of relatively simple interstellar  relevant species (e.g., \ce{H2O}, \ce{NH3}, and \ce{CH3OH}), while radical-radical recombination reactions were thought to be only efficient at temperatures higher than 20--30~K, when radicals are commonly assumed to become mobile \citep[e.g.,][]{Garrod:2013, Butscher:2015}. 

To further strengthen the conclusions of \citet{Fedoseev:2015}, experimental results 
were implemented into a kinetic Monte Carlo model \citep{Cuppen:2009} that was initially built to study CO hydrogenation up to the formation of methanol. This microscopic model showed that in CO-rich ices, HCO radicals can form next to one another through the hydrogenation of CO molecules and react without diffusing in the ice, even for the low H-atom fluxes typical in interstellar space.
The experimental findings were recently complemented by laboratory results on the hydrogen abstraction reactions of \ce{H2CO} and \ce{CH3OH} \citep{Chuang:2016}. Abstraction reactions increase the abundance of radicals in the ice with respect to a purely sequential H-atom addition scheme. This enhances the probability that COMs form through radical-radical recombination without the need of UV photolysis or cosmic rays as external triggers.

We here focus on the latter route: surface reactions through radical-radical recombination at low temperatures and without UV irradiation. We apply a microscopic kinetic Monte Carlo (KMC) model to simulate COM formation at low temperatures for a range of different physical conditions. This will give us constraints on the number of COMs that can be formed through cold grain surface chemistry. Gas-phase routes or cosmic-ray-induced chemistry might contribute as well, but these are beyond the scope of this paper.  This work aims to aid in the interpretation of COM observations by giving insight into how physical conditions affect the obtained abundance ratios. 
The main strength of microscopic simulation methods is that the location of individual species is traced throughout the simulation. An earlier study of hydrogenation of CO \citep{Cuppen:2009} showed that this plays an important role and that a gradient mantle builds up during catastrophic CO freeze-out. This is in agreement with ice observations of CO \citep{Penteado:2015}. In the past years, several codes have been developed that tackle this problem by introducing multiple phases representing different layers within the ice \citep{Taquet:2012,Vasyunin:2013,Furuya:2016}. However, the species are still treated in a ``mean-field'' fashion within a layer, by either rate equations or a macroscopic KMC routine. Crucial to the low-temperature formation mechanism is that two HCO radicals form in close vicinity of each other and can react without the need for diffusion. Obtaining a realistic estimate for the occurrence of this reaction therefore requires knowledge of the exact position of the species with respect to one another. A microscopic KMC model has access to this information and can therefore be used to estimated the relative importance of the different COM formation channels. Moreover, it includes the surface structure, and as our study shows, this has an important effect on the reactivity.

An important starting point for any chemical model is the reaction network, and the accuracy of the simulated data directly depends on the quality of this network. Tunneling reactions and radical-radical reactions play a crucial role in the CO hydrogenation network. In the past years, many tunneling rate constants have been accurately determined, which makes this study very timely. Unfortunately, branching ratios on radical-radical reactions are still lacking. We therefore start this work by determining them.

\section{Method}
\subsection{Electronic structure calculations to determine the branching ratio}\label{sec:branch}
To obtain branching ratios for radical-radical reactions considered in the reaction network, an approach similar to that used in \cite{Lamberts:2018} was employed. In particular, we took the approach to run geometry optimizations for structures where the two radicals are placed in a random orientation with respect to each other. That is, the optimized structure of a radical in the gas phase is placed in the origin and the optimized structure of the second radical is automatically placed next to it at a random angle. This was done 60 times in total, with the C-atom of the second radical being placed symmetrically at the vertices of a truncated icosahedron at a distance of approximately 2.5--4.5 \AA. The initial geometries were then relaxed with an L-BFGS algorithm \citep{Nocedal:1980,Liu:1989} and the optimized geometry indicates the product channel to which a barrierless reaction has lead. The branching ratio of a particular product channel is here defined as its total occurrence devided by 60. Unrestricted density functional theory (DFT) was used with the PW6B95 functional \citep{Zhao:2005} with additional D3 dispersion correction \citep{Grimme:2010} and the def2-TZVP basis set \citep{Weigend:1998}. The ChemShell framework was used for all calculations \citep{Sherwood:2003, Metz:2014}, using Turbomole \citep{Ahlrichs:1989,Treutler:1995} and DL-Find \citep{Kaestner:2009} for the geometry optimizations. For radical-radical reactions it is crucial to use an unrestricted broken-symmetry wave function with overall as many spin-up as spin-down electrons, but finite spin density on both radicals (e.g., \cite{Lamberts:2018,Enrique-Romero:2019}). No Hessian calculations were performed. To test for the influence of the choice of the functional and basis set on the branching ratios, several full geometry optimization runs were repeated using PBEh-3c/mSVP, which is computationally less demanding but includes dispersion correction as well. The major and minor product channels were found to be the same, although the exact values differ, giving credibility to the results. However, we realize that the DFT approach is approximate, and more accurate calculations could be done, particularly with a multi-reference CASSCF approach. 

Although surface reactions on interstellar ices are of interest here, for the current study, a model was employed in which the surface molecules were not explicitly taken into account. The relatively weak interaction between the CO-rich surrounding allow for the assumption that the surface influence on the potential energy landscape of barrierless reactions is small. The D3 dispersion correction can also be seen as mimicking some attractive surface interaction, however. The orientation of the two radicals with respect to each other is important for the outcome \citep{Lamberts:2019}, however. We therefore intentionally used 60 randomly generated starting geometries where the two radicals were placed in close proximity to each other, mimicking a situation in which they are nearest neighbors. With this approach the restricted rotation of a radical because of interactions with hydrogen-bonded \ce{CH3OH}, for example, cannot be taken into account. Sampling a variety of binding sites on realistic dirty ices and determining branching ratios per binding site is beyond the scope of this work, however. 

The initial separation between the two radicals plays a crucial role in whether or not the calculations can reach SCF convergence. When they are too far apart, the interaction will be too small for the radicals to react, whereas when they are too close together, repulsive forces become important that are not representative of the situation on interstellar grains either. We performed some tests for the distance dependence and found that with increasing distance, i) fewer configurations lead to a reaction and more stable nonreactive geometries are calculated instead, and ii) {the branching ratios leading to complex organic molecules (C--C or C--O bonds) are lower because COM formation is more sterically hindered in general.} For the current study, we mimicked the situation that two radicals are on the surface in close proximity (nearest neighbors), as is the case for the KMC simulations as well, with a closest atom distance of 2.5--4.5~{\AA} between the two radicals at hand. This separation depends on the size of the reactants themselves and is comparable to the site-site distance in the KMC model.

\subsection{Kinetic Monte Carlo simulations}
Ice evolution was simulated using a continuous-time, random-walk (CTRW) algorithm, which is a type of KMC method. For an in-depth theoretical description of this technique, see \citet{Cuppen:ChemRev}. The program used in this study is similar to the one used in \citet{Cuppen:2007A} and \citet{Cuppen:2009}. Briefly, the KMC technique simulates the evolution of a grain surface. This grain surface is represented by a $50\times 50$ grid with periodic boundary conditions and is bare at the start of the simulation. A random-number generator in combination with the rates determines the sequence of processes. Similar to \citet{Cuppen:2009}, the types of processes include deposition onto the surface, reaction of two species, hopping between sites, and desorption from the surface. These processes have been extended to also include photodissociation and cosmic-ray dissociation processes \citep{Cuppen:2007A}. It is assumed that the rate coefficients for hopping and desorption are thermally activated by
\begin{equation}
        R_\text{X} = \nu \exp\left(-\dfrac{E_\text{X}}{T}\right)
,\end{equation}
where $\nu$, the preexponential factor, has been approximated by $kT/h\approx 2\cdot10^{11} \text{ s}^{-1}$ and $E_\text{X}$ is the energy barrier of process X in K. The rate coefficients for the photodissociation and cosmic-ray dissociation processes are determined according to
\begin{equation}
        R_{\text{photo}}=\alpha_{\text{photo}} \exp\left(-\gamma_{\text{photo}}A_\text{V}\right)
\end{equation}
and
\begin{equation}
        R_{\text{CR}}=\alpha_{\text{CR}}\zeta,
\end{equation}
respectively. Here $\alpha$ is the photon rate, $\gamma$ is the extinction coefficient, $A_\text{V}$ is the visual extinction, and $\zeta=1.3\cdot10^{-10} \text{ yr}^{-1}$ is the cosmic-ray ionization rate. Standard values from the UMIST RATE2012 database were taken \citep{McElroy:2013}.

The program was used to simulate laboratory conditions and interstellar conditions. The former was used as a benchmark, the latter to obtain insight into the conditions under which complex organic species are formed. A full overview of all performed simulations can be found in Table~\ref{tab:simulations}.

\begin{table}
 \caption{Overview of all simulations.}
 \label{tab:simulations}
 \centering
 \begin{tabular}{l r r l r}
  \hline\hline
                & $T$ [K]       & $n(\ce{H})$ [cm$^{-3}$]       & $n(\ce{CO})$ [cm$^{-3}$]     & $A_\text{V}$ [mag]    \\
  \hline
        \multicolumn{5}{c}{Experimental}                                \\
        1       & 14            & 1.18E+08      & 2.53E+08      & 20                            \\
        2       & 15            & 7.10E+07      & 2.63E+08      & 20                            \\
        3       & 15            & 7.10E+07      & 2.72E+08$^\dagger$    & 20      \\
  \hline
        \multicolumn{5}{c}{Experimental (UMIST)}                                                \\
        4       & 14            & 1.18E+08      & 2.53E+08      & 20                            \\
        5       & 15            & 7.10E+07      & 2.63E+08      & 20                            \\
        6       & 15            & 7.10E+07      & 2.72E+08$^\dagger$    & 20      \\
  \hline
        \multicolumn{5}{c}{Interstellar}                                \\
        7 A--D  &  8    & 0.5--6.0      & 10    & 20    \\
        8 A--D  & 10    & 0.5--6.0      & 10    & 20    \\
        9 A--D  & 12    & 0.5--6.0      & 10    & 20    \\
        10 A--D & 14    & 0.5--6.0      & 10    & 20    \\
        11 A--D & 16    & 0.5--6.0      & 10    & 20    \\
        12 A--D & 18    & 0.5--6.0      & 10    & 20    \\
        13 A--D & 20    & 0.5--6.0      & 10    & 20    \\
  \hline\hline
  \multicolumn{5}{l}{\footnotesize $\dagger$: \ce{H2CO} instead of \ce{CO}}

  \end{tabular}
 
\end{table}

\section{Results and discussion}

\subsection{Reaction network}
\label{sec:reaction_network}
 
The reaction network used in this study is given in Table~\ref{tab:ratestable}. The same color-coding scheme for the main species is used in all figures throughout the paper. The basis of this reaction network is a hydrogen addition chain from carbon monoxide (yellow) through formalaldehyde (orange) to methanol (red) \citep{Watanabe:2002,Fuchs:2009}. When the lifetime of formyl (\ce{HCO}) radicals is sufficient, two of these radicals could recombine to form glyoxal (\ce{HC(O)CHO}, GX) \citep{Fedoseev:2015,Chuang:2016}. Through continued hydrogenation, glyoxal will consecutively be turned into glycolaldehyde (\ce{HC(O)CH2OH}, GA, purple) and ethylene glycol (\ce{H2C(OH)CH2OH}, EG, green). Finally, recombination of a formyl and a methoxy radical leads to the formation of methyl formate (\ce{HCOOCH3}, MF, blue). 

Two types of reactions are present in the reaction network. First, recombination reactions involving two radical species. These reactions are assumed to be barrierless and are included with a reaction rate $k = kT/h \approx 2\cdot 10^{11} \text{s}^{-1}$ at these low temperatures. Second, hydrogenation reactions that possess barriers. The reaction pathways of such reactions have been calculated by \citet{Andersson:2011}, \citet{Song:2017}, and \citet{Alvarez-Barcia:2018} using instanton theory, and they were optimized through a quasi-Newton-Rhapson method. A resulting instanton is the most probable tunneling path for a given reaction. The corresponding tunneling rate can be calculated using harmonic quantum transition state theory (HQTST).  The reaction rate is governed by tunneling below the critical temperature $T_\text{c}$ , which is
\begin{equation}
        T_\text{c} = \frac{\hbar\Omega}{2\pi k_B},
\end{equation}
where $\Omega$ is the imaginary normal-mode frequency of the saddle point. The crossover temperature is equal to 79~K for a typical reaction \ce{H + CO -> HCO} with a frequency of 2171~cm$^{-1}$\citep{Andersson:2011}.

An important addition to the reaction network with respect to previous versions \citep{Cuppen:2009,Fedoseev:2015} is the inclusion of abstraction reactions such as \ce{H2CO + H -> HCO + H2}. \citet{Penteado:2017} have shown that competition between this abstraction and the two hydrogenation reactions toward \ce{CH3O} and \ce{CH2OH} has a definitive influence on the yields of the species. \citet{Chuang:2016} have already shown that hydrogen-addition to formaldehyde yields significantly less \ce{CH2OH} than either \ce{CH3O} or \ce{HCO}, which has been confirmed by the calculations of \citet{Song:2017}.

\subsubsection{Branching ratios of radical-radical reactions}
The radical-radical branching ratios  were estimated using the method described in Section~\ref{sec:branch}. 
The branching ratios obtained here for a consistent set of reactions are the first to be reported in the literature and serve as a guide for the major and minor product channels. Product channels with branching ratios below 0.05 ($<$5\%) are listed only in Table~\ref{tab:ratestable}, but are not included in the KMC model. 
We did not consider any reactions between radicals of the sort \ce{C2H_nO2} and HCO, \ce{CH2OH}, or \ce{CH3O}. Although we are aware that these reactions can take place in a barrierless fashion as well (see also \cite{Butscher:2017}), they are beyond the scope of this study.

The branching ratios in Table~\ref{tab:ratestable} do not add up to one for all reactions. The reactants \ce{HCO + HCO} are an example. This is because certain starting geometries did not lead to reaction even though the reaction being barrierless in principle. This is most likely because their relative geometry does not lead to a suitable reaction complex that leads to reaction, and hence no reaction occurs. 
This effect could also occur on a grain surface at low temperatures where two species meet in the wrong geometry and cannot reorient. The two species then remain frozen in close proximity. The KMC program was changed to allow for nonreactive branches to accommodate for this effect. This was realized by first choosing which reaction was added to the table of events in the KMC routine. This choice was based on the branching ratios, and in principle, also no reaction can be added, in which case the species were only allowed to diffuse or desorb. Simply lowering the overall reaction rate will not help to account for this because it is a stochastic process that cannot be repaired in a mean-field way, that is,\emph{} smearing out the lowered reactions. For HCO recombination, for instance, there is a chance of 0.40 that the reaction \ce{2HCO ->  HCOCHO} with rate $2 \cdot 10^{11}$~s$^{-1}$ is added to the table of events, a chance of 0.25 for \ce{2HCO ->  H2CO + CO} with rate $2 \cdot 10^{11}$~s$^{-1}$, and 0.35 that no reactions are added. We expect this effect to become increasingly important for larger reactants. We show below that this inclusion does not significantly change the results.

\renewcommand{\baselinestretch}{1}
\begin{table*}[ht!]
\centering
\caption{Overview of all reaction rates}
\label{tab:ratestable}
\begin{tabular}{l@{$\rightarrow$ }l r r r r}
\hline\hline
\multicolumn{2}{l}{Reactions$^\dagger$} & \multicolumn{2}{l}{Current}   & \multicolumn{2}{l}{UMIST}       \\
\multicolumn{2}{l}{}                                    & Rate$^\ddagger$ [s$^{-1}$]      & Branch        & Rate [s$^{-1}$]       & Branch        \\
\hline
\ce{H}                  + \ce{H}                & \ce{H2}                                       & 2.00E+11        & 1.00E+00              & 2.00E+11      & 1.00E+00      \\
\ce{CO}                 + \ce{H}                & \ce{HCO}                                      & 2.00E+05        & 1.00E+00              & 2.50E+03      & 1.00E+00      \\
\ce{HCO}                + \ce{H}                & \ce{H2CO}                                     & 2.00E+11        & 4.17E$-$01    & 2.00E+11      & 1.00E+00      \\
                                                                & \ce{CO}               + \ce{H2} & 2.00E+11      & 5.83E$-$01    &                       &                       \\
\ce{H2CO}               + \ce{H}                & \ce{CH2OH}                            & 1.50E+05        & 7.27E$-$05    & 2.50E+03      & 3.33E$-$01\\
                                                                & \ce{CH3O}                                     & 1.50E+05        & 3.33E$-$01    & 2.50E+03      & 3.33E$-$01\\                                                          & \ce{HCO}                + \ce{H2}       & 1.50E+05      & 3.33E$-$01    & 2.50E+03        & 2.44E$-$01\\
\ce{CH2OH}              + \ce{H}                & \ce{CH3OH}                            & 2.00E+11        & 6.00E$-$01    & 2.00E+11      & 1.00E+00      \\
                                                                & \ce{H2CO}             + \ce{H2} & 2.00E+11      & $<$5.00E$-$02 &                       &                       \\
                                                                & \ce{H2COH2}                           & 2.00E+11        & $<$5.00E$-$02 &                       &                       \\
\ce{CH3O}               + \ce{H}                & \ce{CH3OH}                            & 2.00E+11        & 3.00E$-$01    & 2.00E+11      & 1.00E+00      \\
                                                                & \ce{H2CO}             + \ce{H2} & 2.00E+11      & 1.50E$-$01    &                       &                       \\
\ce{CH3OH}              + \ce{H}                & \ce{CH3O}             + \ce{H2} & 2.00E+01      & 5.00E$-$01    &                       &                       \\
                                                                & \ce{CH2OH}    + \ce{H2} & 4.00E+02      & 5.00E$-$01    &                       &                       \\
\hline
\ce{HCO}                + \ce{HCO}              & \ce{HCOCHO}                           & 2.00E+11        & 4.00E$-$01    &                       &                       \\                                                              & \ce{H2CO}               + \ce{CO}       & 2.00E+11      & 2.50E$-$01    &                       &                       \\
                                                                & \ce{t-HCOH}   + \ce{CO} & 2.00E+11      & $<$5.00E$-$02 &                       &                       \\
                                                                & \ce{CHOHCO}                           & 2.00E+11        & $<$5.00E$-$02 &                       &                       \\
\ce{HCO}                + \ce{CH2OH}    & \ce{YCHO}                                     & 2.00E+11        & 3.50E$-$01    & 2.00E+11      & 1.00E+00      \\
                                                                & \ce{H2CO}             + \ce{H2CO}       & 2.00E+11      & 5.00E$-$02    &                       &                       \\
                                                                & \ce{CH3OH}    + \ce{CO} & 2.00E+11      & 4.00E$-$01    &                       &                       \\
\ce{HCO}                + \ce{CH3O}             & \ce{HCOOCH3}                          & 2.00E+11        & 2.50E$-$01    & 2.00E+11      & 1.00E+00      \\
                                                                & \ce{H2CO}             + \ce{H2CO}       & 2.00E+11      & 2.00E$-$01    &                       &                       \\
                                                                & \ce{CO}               + \ce{CH3OH}& 2.00E+11    & 4.00E$-$01    &                       &                       \\
\ce{H2CO}               + \ce{CH3O}             & \ce{HCO}              + \ce{CH3OH}& 2.00E+03    & 4.00E$-$01    & 1.50E+03      & 5.00E$-$01\\
\ce{CH2OH}              + \ce{CH2OH}    & \ce{YCH2OH}                           & 2.00E+11        & 4.00E$-$01    &                       &                       \\
\ce{CH2OH}              + \ce{CH3O}             & \ce{CH3OH}    + \ce{H2CO}     & 1.00E+11        & 7.20E$-$01    &                       &                       \\
\ce{CH3O}               + \ce{CH3O}             & \ce{CH3OH}    + \ce{H2CO}     & 1.00E+11        & 1.00E+01              &                       &                       \\
\hline
\ce{HCOCHO}             + \ce{H}                & \ce{H2COCHO}                          & 9.60E+06        & 3.33E$-$01    &                       &                       \\
                                                                & \ce{HCOHCHO}                          & 9.60E+06        & 3.33E$-$05    &                       &                       \\
                                                                & \ce{COCHO}    + \ce{H2} & 9.60E+06      & 3.33E$-$01    &                       &                       \\
\ce{H2COCHO}    + \ce{H}                & \ce{YCHO}                                     & 2.00E+11        & 5.00E$-$01    &                       &                       \\\ce{HCOHCHO}  + \ce{H}          & \ce{YCHO}                                     & 2.00E+11      & 5.00E$-$01      &                       &                       \\\ce{YCO}              + \ce{H}          & \ce{YCHO}                                     & 2.00E+11      & 5.00E$-$01      &                       &                       \\
\ce{YCHO}               + \ce{H}                & \ce{YCH2O}                            & 2.80E+05        & 4.00E$-$01    &                       &                       \\
                                                                & \ce{YCHOH}                            & 2.80E+02        & 1.00E$-$01    &                       &                       \\
                                                                & \ce{YCO}              + \ce{H2} & 7.80E+07      & 4.00E$-$01    & 2.40E+03      & 5.00E$-$01\\
                                                                & \ce{HCOHCHO}  + \ce{H2} & 2.60E+04      & 1.00E$-$01    &                       &                       \\
                                                                & \ce{H2COCHO}  + \ce{H2} & 5.60E+05      & 4.60E$-$02    &                       &                       \\
                                                                & \ce{CH3OH}    + \ce{HCO}        & 5.60E+05      & 1.70E$-$06    & 2.40E+03      & 5.00E$-$01\\
\ce{YCHO}               + \ce{CH3O}             & \ce{HCOOCH3}  + \ce{CH2OH}&                   &                               & 2.40E+03        & 5.00E$-$01\\
\ce{YCH2O}              + \ce{H}                & \ce{YCH2OH}                           & 2.00E+11        & 5.00E$-$01    &                       &                       \\\ce{YCHOH}            + \ce{H}          & \ce{YCH2OH}                           & 2.00E+11      & 5.00E$-$01      &                       &                       \\\ce{YCH2OH}           + \ce{H}          & \ce{YCHOH}    + \ce{H2}       & 3.50E+06      & 5.00E$-$01    &                       &                       \\
                                                                & \ce{YCH2O}    + \ce{H2} & 3.50E+06      & 5.00E$-$04    &                       &                       \\
\ce{HCOOCH3}    + \ce{H}                & \ce{CH3OH}    + \ce{HCO}      & 3.80E$-$33& 1.00E+00            & 2.45E+03      & 5.00E$-$01\\
\hline\hline
\multicolumn{6}{l}{\footnotesize$\dagger$: For readibility of the table, a fully hydrogenized \ce{CO} group, i.e., \ce{CH2OH,} is represented by Y. }\\
\multicolumn{6}{l}{$\ddagger$: Radical reactions have been assumed to be barrierless with a reaction rate of $2\cdot10^{11}$~s$^{-1}$.}
\end{tabular}
\end{table*}
\renewcommand{\baselinestretch}{1.5}

\subsubsection{Activated reactions}
\cite{Alvarez-Barcia:2018} calculated rate constants for GX + H leading to \ce{H2COCHO} and \ce{HCOHCHO}. They were unable to treat the abstraction reaction because their quantum chemical method could not accurately describe this particular reaction. Gas-phase experiments at room temperature show, however, that this is an important reaction that leads to HCOCO, which can spontaneously decay to CO + HCO \citep{Colberg:2006,Orlando:2001}. This two-step reaction is included in the network as the main destruction route of the C--C bond. We assumed the rate constant to be comparable to its hydrogenation. Room-temperature experimental data show a rate constant of roughly three orders of magnitude higher than abstraction of \ce{H2CO}. For the unimolecular dissociation, we used a rate constant of $10^4$~s$^{-1}$ , which is a factor of 600 lower than the rate constant at 250~K \citep{Orlando:2001}. With this rate, the entire HCOCO decays to CO and HCO. This is rather insensitive to the exact value.

The only radical-neutral reaction with a radical other than the H atom that we considered is \ce{H2CO + CH3O -> CH3OH + HCO}. Although \cite{Butscher:2017} indicated that the reaction between the HCO radical and \ce{H2CO} might also take place, in fact \cite{Alvarez-Barcia:2018} showed that reactions where heavy atoms play a role in the bond formation are not significantly enhanced by tunneling. In other words, the rate constants for the reactions \ce{H2CO + CH3O -> CH3OCH2O} and \ce{H2CO + HCO -> (HCO)CH2O} are below $10^{-9}$ s$^{-1}$ at low temperatures. Therefore other reactions of the same type, such as the formation of a C--C or C--O bond through radical-neutral reactions, were discarded as well. The reaction \ce{HCO + CO -> CO + HCO} may lead to transport of H atoms in the CO bulk. This reaction is expected to have a high barrier, however, because a C--H bond needs to be broken and the reaction \ce{H + CO -> HCO} already has a high barrier itself.

In principle, branching ratios for different channels in activated reactions can be determined by simply equating them to the relative rate constants for the individual reactions as determined quantum chemically. However, this is an oversimplification and might lead to ignoring important product channels. For the reaction GA + H, this can result in four product channels of which hydrogen addition to 
\ce{CH2OHCH2O} ($k_1 = 2.8 \times 10^5$~s$^{-1}$) and abstraction leading to \ce{HCOHCHO} ($k_2 = 6.8 \times 10^7$~s$^{-1}$) are the major outcomes. The ratio between the rate constants $k_2/k_1$ suggests that abstraction is 200 times more likely than addition. On a grain surface this is not the case, however. When a hydrogen atom approaches GA, the possible product channel will depend on the direction of approach, whether this reaction will indeed occur depends on the residence time of the H atom at close vicinity and on possible roaming of the H atom during this time. Considering diffusion rates of hydrogen atoms, which are much slower than both reaction rates, we expect both reactions to occur when meeting in the correct geometry for that particular reaction. The branching ratios of both channels will be much more comparable than based on their relative rates. We therefore chose to treat reactions with multiple channels in a similar way as the nonactivated reactions: a reaction is chosen to be added to the table of events with more or less equal probability (see branching ratio column in Table~\ref{tab:ratestable}). This reaction is added with its corresponding rate. If GA and H are in adjacent sites, there is a probability of 0.40 that \ce{GA + H -> CH2OHCH2O} with rate $k_1$ is added to the table of events and a probability of 0.40 for \ce{GA + H -> HCOHCHO + H2} with rate $k_2$. This leads to a roughly equal occurrence of both reactions.

\subsection{Experimental simulations}

The reaction network was benchmarked against the experimental results as published by \citet{Chuang:2016,Chuang:2017}. Table~\ref{tab:koju} shows an overview of their results as they  are relevant to our study. All three experiments are codeposition experiments of either CO and H or \ce{H2CO} and H over 6 hours. The surface abundances of CO, \ce{H2CO,} and \ce{CH3OH} were followed by reflection absorption infrared spectroscopy (RAIRS) \emph{\textup{in situ}} during the experiments, whereas relative abundances of MF, GA, and EG were only obtained after heating the sample at the end of the experiment. The abundances of CO, \ce{H2CO}, and \ce{CH3OH} in Table~\ref{tab:koju} are shown in monolayers (MLs). For the COMs with C--C bond, the final ratios for MF:GA:EG are given. In the benchmark study, the simulations ran for an equivalent time of 6~hours or 21600~seconds, at which time a steady state was reached in most cases, and the overall surface coverage was then compared against the experimental results. We used an effective H-atom flux of $f_\text{effective} = f_\text{exp} \times S, $ where $S$ is the sticking fraction of H on a surface, which we assumed to be 0.20 \citep{Buch:1991,Al-Halabi:2002,Al-Halabi:2003,Al-Halabi:2004,Batista:2005,Veeraghattam:2014} for a room-temperature atom landing on a cold surface. Reducing the sticking fraction from unity to the more realistic value of 0.2 leads to a slightly better agreement. 

As mentioned earlier, the majority of our reaction rate constants are based on \citet{Alvarez-Barcia:2018}. One of the most critical reactions to obtain agreement with the experiments is \ce{H + H2CO}. This reaction with three possible product channels has been studied quantum-chemically by \citet{Song:2017}. They studied this reaction both in the gas phase and on a water-ice (ASW) surface. The ASW changes the transition state for these reactions through hydrogen bonding, which results in significantly higher rates than in the gas phase. Because we work on a CO surface without hydrogen-bonding capabilities, we expect the current system to be best represented by gas-phase conditions. When we use these rates directly, \ce{H2CO} is predominantly destroyed to form HCO and \ce{H2}. Both methanol and formaldehyde are highly underproduced. The actual rate constants calculated by \citet{Song:2017} are based on DFT calculations. The applied functional is first compared against stationary calculations at CCSD(T)-F12 level, which is more accurate, but also much more computationally demanding. This comparison shows that the reaction barriers are underestimated, and the reaction rate constants are accordingly overestimated. This effect is stronger for the abstraction reaction than for the hydrogenation reaction leading to \ce{CH3O}. Decreasing the rates and using equal branching ratios leads to much better results that are within the uncertainty of the method.

\begin{table*}[tb!]
        \caption{Summary of the experimental results as published by \citeauthor{Chuang:2016}}
        \label{tab:koju}
        \centering
        \begin{tabular}{l l l l r r r r@{ : }r@{ : }l r}
                \hline
                \hline
                                & Experiment    &       Ratio   &       $T$ [K]     &       \multicolumn{3}{c}{Coverage (MLs)}                      &       \multicolumn{3}{c}{Relative abundance}      & ref\\
                                &                               &                       &                       &       \ce{CO}                 &       \ce{H2CO}               &       \ce{CH3OH}      &        MF  & GA & EG                           &\\
                \hline
                1               & \ce{H + CO}   &       16:1    &       14              &       6.70                    &       4.19                    &       1.67            &        0.3 & 64 & 36                           & a\\
                2               & \ce{H + CO}   &       14:1    &       15              &       10.51                   &       2.75                    &       0.59            &        0.0 & 59 & 41                           & b\\
                3               & \ce{H + H2CO} &       14:1    &       15              &       0.95                    &       10.01                   &       2.27            &        5.8 & 53 & 42                           & b\\
                \hline
                \hline
        
        \multicolumn{4}{l}{\footnotesize a: \citet{Chuang:2017}} \\
        \multicolumn{4}{l}{\footnotesize b: \citet{Chuang:2016}} \\
        \end{tabular}
        
\end{table*}

\begin{figure*}
 \includegraphics[width=\textwidth]{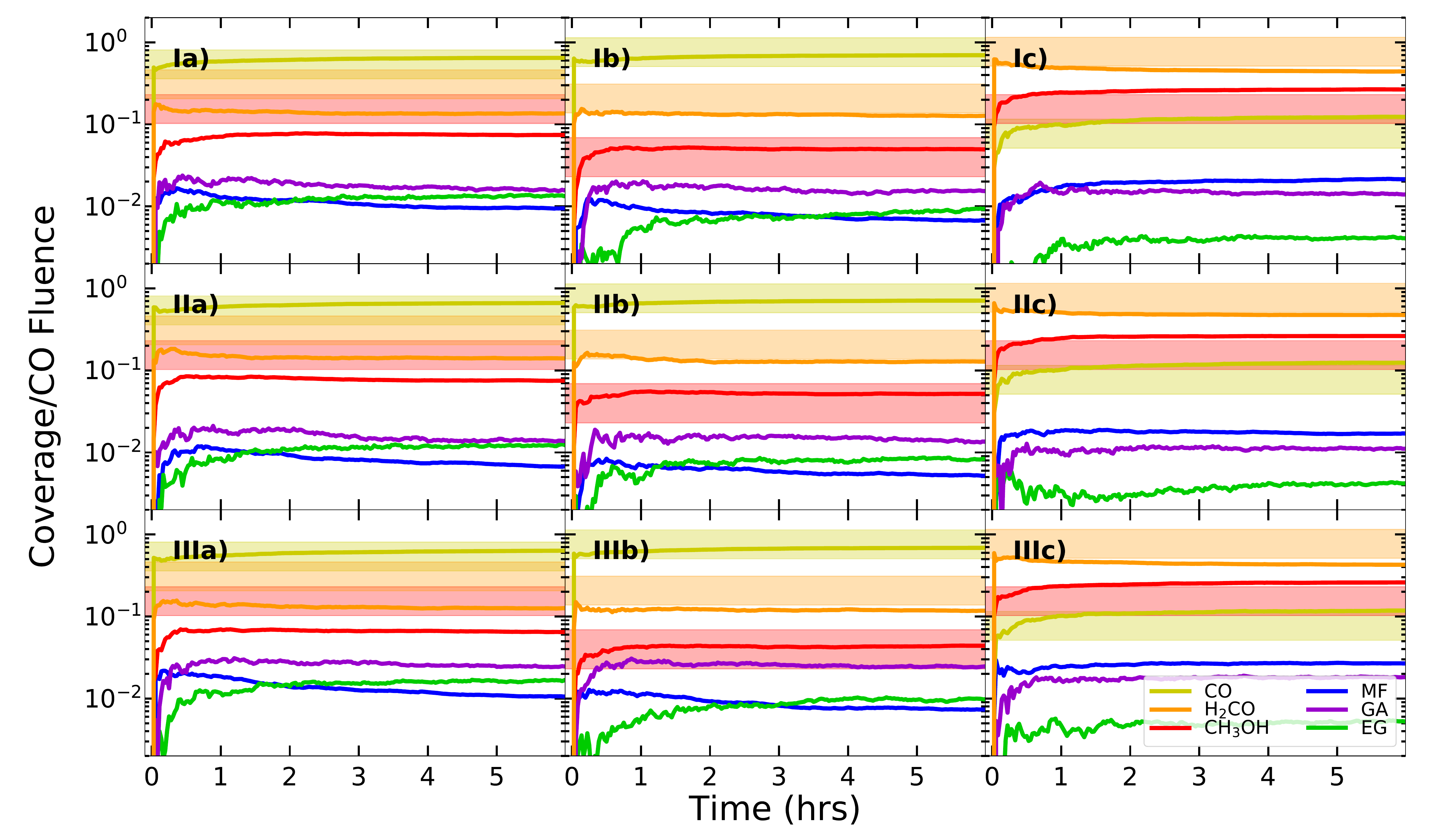}
        \caption{Simulated codeposition of CO+H (a and b) and \ce{H2CO}+H (c) or simulations 1-3 of Table~\ref{tab:simulations}. The three different rows correspond to different choices of H-atom binding energy  (see main text). I: 250~K, II: 420~K, and III: 670~K.}
        \label{fig:all_exp}
\end{figure*}

Figure~\ref{fig:all_exp} shows the simulated build-up during codeposition of H and CO or \ce{H2CO}. The \ce{H} input flux of simulation 1 (Fig.~\ref{fig:all_exp}a) is equal to $1.6\cdot10^{12}$~cm$^{-2}$ and the flux of the codeposited species equals $6.5\cdot10^{11}$~cm$^{-2}$s$^{-1}$. For simulations 2 and 3 (Fig.~\ref{fig:all_exp}b,c), these fluxes are equal to $1.4\cdot10^{13}$ and $7.0\cdot10^{11}$~cm$^{-2}$s$^{-1}$ , respectively. In all simulations, a molecular hydrogen flux was added equal to that of the atomic hydrogen flux to account for recombination in the cooling nozzle of the particle beam. Along the $y$-axis, the graphs show the deposition of several species relative to the CO fluence, for example, the total number of monolayers of deposited CO. The colored translucent bands indicate the results obtained by \citet{Chuang:2016}, maintaining an uncertainty factor of $1.5$. The three different rows correspond to three different sets of H-atom binding energy. We used a fixed diffusion barrier for hydrogen on a flat CO surface of 200~K. Diffusion barriers are site-specific and depend on the binding energy before and after the hop, $\Delta E$: $E_\text{diff} = E_\text{diff}^\text{flat} + \max(\Delta E,0)$~K. The value of 200~K is in line with experimental results by \citet{Kimura:2018} of H diffusion on CO ice and agrees well with calculations of long-range H diffusion on ASW by \citet{Asgeirsson:2017} and \citet{Senevirathne:2017}. The binding energy of H to a CO surface is less well constrained, however. To test the sensitivity of the results to the H-binding energy, we varied this. Panels I correspond to a binding energy of 250~K (flat surface), panels II to 420~K, and  panels III to 670~K.  Small differences can be observed between the various plots, but the final abundance is rather independent of the choice of binding energy. Experiment 2 is quite well reproduced by the model, whereas the model slightly overproduces \ce{CH3OH} for experiment  3 and slightly underproduces it for experiment 1. Even when we varied the rates, we were not able to arrive at a model that could reproduce all experimental results simultaneously. In the remainder of the paper, the intermediate value of $E_\text{bind}(\ce{H2}) = 420$~K is used.

Table~\ref{tab:codeposition} summarizes the abundances at the end of simulations of Fig.~\ref{fig:all_exp} II and can be directly compared to Table~\ref{tab:koju}.  It is quite a triumph for the computational chemistry that the model that uses purely calculated rate constants and branching ratios can reproduce the experimental results this well. The main discrepancy is the formation of MF. The formation trends are the same between experiment and simulation, but MF is much more abundant in the simulations than in the experiments. In our model, MF is formed through \ce{HCO + CH3O -> HCOOCH3} , and once it is formed, there is no destruction reaction. The overabundance in our model is hence either due to an overestimation of the branching ratio of \ce{HCO + CH3O -> HCOOCH3} or due to missing reactions. From quantum chemistry calculations, we know that abstraction reactions such as \ce{HCOOCH3 + H -> COOCH3 + H2} have very low rates \citep{Alvarez-Barcia:2018}. Other destruction reactions are not obvious based on the species available in our system.

\begin{figure}
 \includegraphics[width=0.45\textwidth]{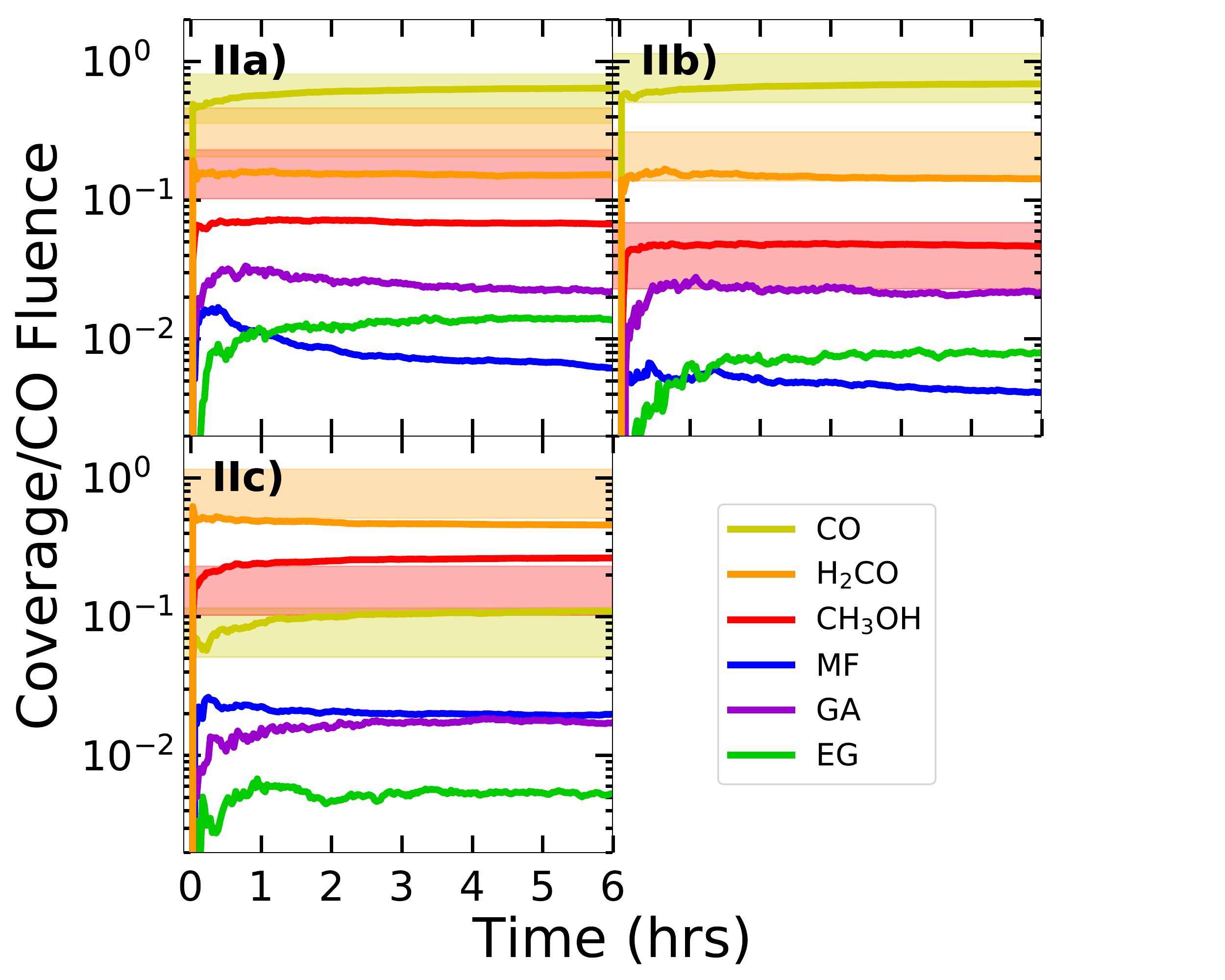}
 \caption{Same as for Fig.~\ref{fig:all_exp}II, but without nonreactive branches.}
 \label{fig:no-non-rea}
\end{figure}

The removal of nonreactive branches has only a slight effect on the final results, as can be seen from Fig.~\ref{fig:no-non-rea} and comparing it to Fig.~\ref{fig:all_exp}II for the same conditions. When all radical-radical encounters lead to reaction, slightly more glycolaldehyde is formed at the expense of \ce{CH3OH}, but other species remain rather unaffected. Whether two radicals can indeed remain frozen next to each other in the ice depends on their reorientation timescale, which is unknown. The present system is a CO-rich ice where hydrogen bonding plays a minor role and species might reorient. On a water-ice surface, we expect species with hydrogen-bonding capabilities to be more easily frozen into a specific geometry, and nonreactive branches can then become important in the grain surface chemistry, especially for larger species. How nonreactive branches should be considered is a topic for further investigation. To present a conservative lower limit of the COM abundances, we included nonreactive branches in our simulations of ISM conditions.

\begin{table}
        \caption{Obtained surface abundances at the end of the simulated experiments.}
        \centering
 \begin{tabular}{l r r r r r r r}
        \label{tab:codeposition}\\
        \hline
        \hline
                        & $T$   (K)     & \multicolumn{6}{c}{Coverage (MLs)}    \\
                        &               & CO    & \ce{H2CO}     & \ce{CH3OH}    & MF      & GA    & EG    \\
        \hline
        1               & 14    & 10.69 & 2.25          & 1.23  & 0.16  & 0.26    & 0.22  \\
        2               & 15    & 10.01 & 1.83          & 0.71  & 0.10  & 0.22    & 0.13  \\
        3               & 15    &  2.02 & 7.25          & 4.37  & 0.35  & 0.23    & 0.07  \\
        \hline
        \hline
 \end{tabular}
\end{table}
\subsection{Comparison with UMIST rates}

\begin{figure}
 \includegraphics[width=0.45\textwidth]{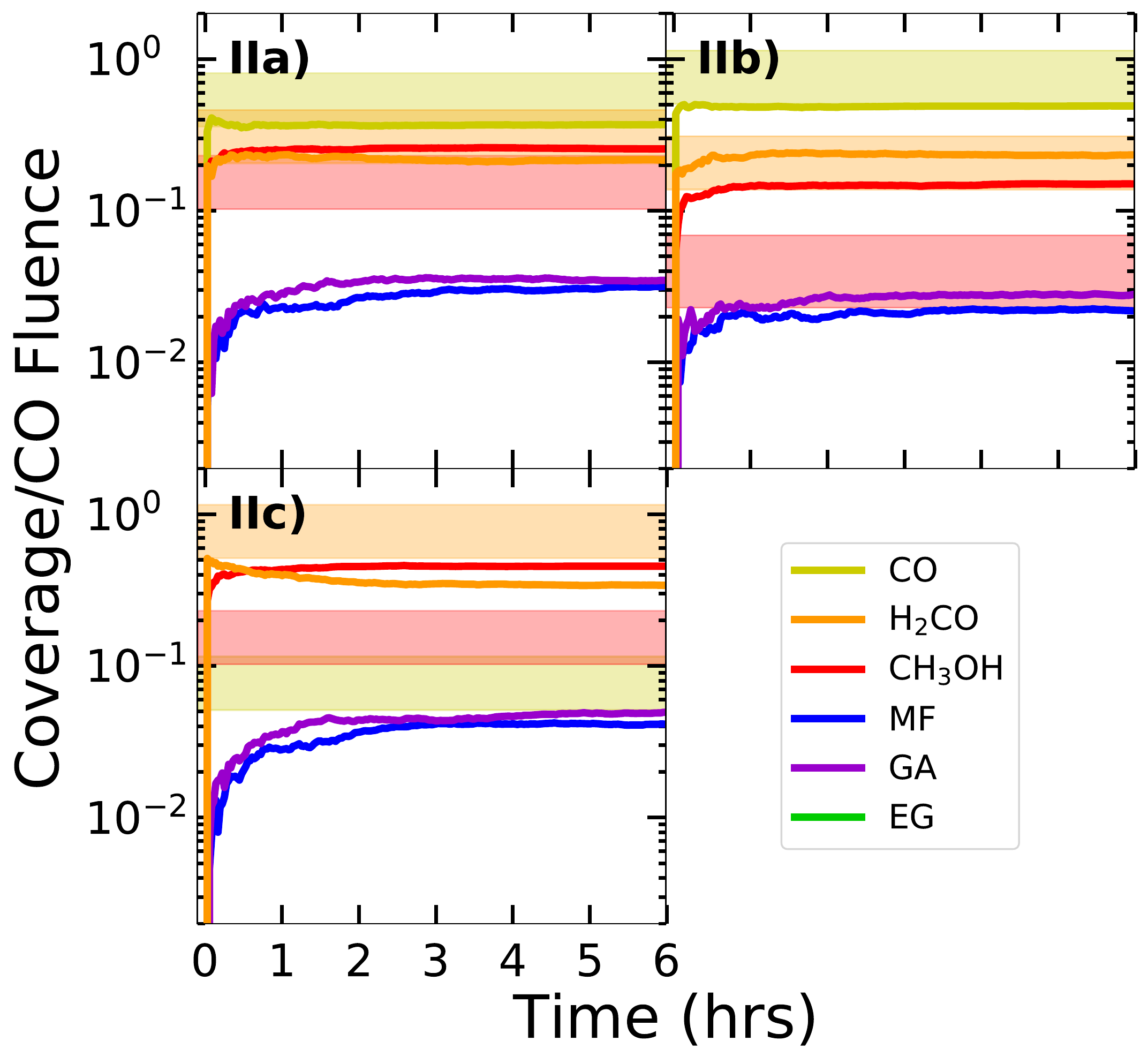}
    \caption{Same as Fig.~\ref{fig:all_exp}II  using UMIST as a source for reaction rate constants instead of \citet{Alvarez-Barcia:2018} (Table.~\ref{tab:ratestable})}
        \label{fig:UMIST}
\end{figure}

The UMIST RATE2012 database (\emph{see: astrochemistry.net} \citep{McElroy:2013}) is generally used as a source of rate constants for astrochemical models. Our new network, which is based on a consistent set of calculated tunneling rates and is benchmarked against experiments, is a substantial improvement over the UMIST values for this network. The large number of missing reactions in the latter network is immediately apparent when the new rates are compared with the UMIST rates in Table~\ref{tab:ratestable} . This involves mostly abstraction reactions toward less complex species and the HCO + HCO reactions and the subsequent hydrogenation reactions toward GA and EG. The second point that stands out is that the rates for \ce{H + CO} and \ce{H + H2CO} are two orders of magnitude higher in the new network. The low-rate hydrogenation reactions in the UMIST database are hence an effective rate for the full hydrogenation chain. We would also like to point out the reaction constant of hydrogenation of MF, which is 36 orders of magnitude lower according to \citet{Alvarez-Barcia:2018} than the constant included in UMIST. Rate constants in UMIST are typically calculated based on a barrier height, assumed barrier width, and an estimated effective mass using a crude formula for tunneling. The effective mass estimation completely fails for this reaction because it does not only involve movement of H, but also breaking a C--O bond.

We performed the same set of three simulations using rates from the UMIST network (Table~\ref{tab:ratestable}). 
The results of these simulations are shown in Fig.~\ref{fig:UMIST}. Whereas the agreement for the experiment a is quite reasonable, methanol is heavily overproduced for b and c. The reason for this is that the UMIST network uses a lower effective rate instead of the full hydrogenation or abstraction network. As long as sufficient H atoms are available, CO is fully hydrogenated toward \ce{CH3OH}. When methanol is formed, no destruction pathway leadis away from this. The dependence of the simulated coverages on the chosen sticking fraction of the H atoms to the surface is also much larger than for the new network.  The need of an abstraction route is further apparent in the \ce{H2CO}-hydrogenation simulation (panel c), which shows no CO at all.  
Only MF and GA are formed of the two-carbon COMs. Their abundance is very similar, whereas experiments show that MF is much less produced than GA, and their ratio also differs between the different experimental conditions.

Based on these simulations, we therefore expect the UMIST network to predict a too high methanol and MF abundance in dark clouds and a too low abundance of GA and EG. This would automatically favor the photodissociation heat-up mechanism by \citet{Garrod:2006a}.


\subsection{Interstellar conditions}
The second part of this study aims to use the benchmarked network to simulate the evolution of a grain-surface mantle in dark molecular cloud conditions. In the ISM simulations, the flux of CO and H onto the grain surface depends on the gas-phase abundance of both species. For H atoms, we adopted a constant gas-phase abundance throughout the simulation. Because the CO abundance gradually decreases as more CO is depleted, the CO gas-phase abundance was recalculated during the simulation, taking the current surface CO and its derivatives into account. Because molecular hydrogen is nearly inert in our reaction network (see Section \ref{sec:reaction_network}), we kept it at a constant low abundance of 10~cm$^{-3}$ for all simulations to save computational time. A subset of the simulations was repeated with a higher \ce{H2} abundance of 100~cm$^{-3}$ , and we found no significant difference. 

\begin{figure}[tb!]
        \includegraphics[width=0.45\textwidth]{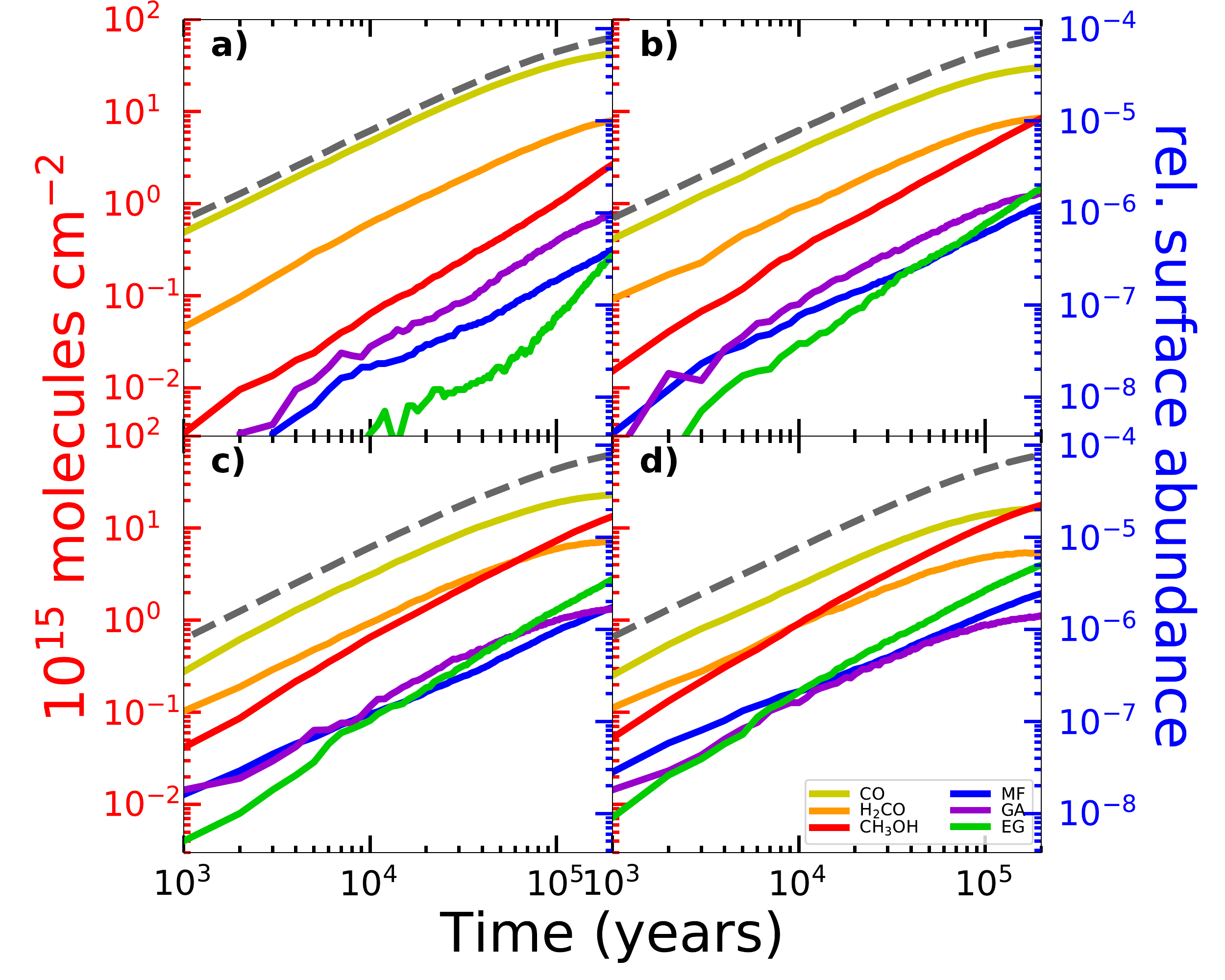}
        \caption{Simulated time evolution of the grain surface composition for four different levels of hydrogen abundances: 1.0, 2.5, 4.0, and 6.0~cm$^{-3}$ (a-d respectively). The abundance of gas-phase carbon monoxide is 10.0~cm$^{-3}$. The temperature of the grain is 10~K. The relative surface abundance is given with respect to $n_\text{H}$.}
        \label{fig:ISM_time}
\end{figure}

\begin{figure}[tb!]
        \includegraphics[width=0.45\textwidth]{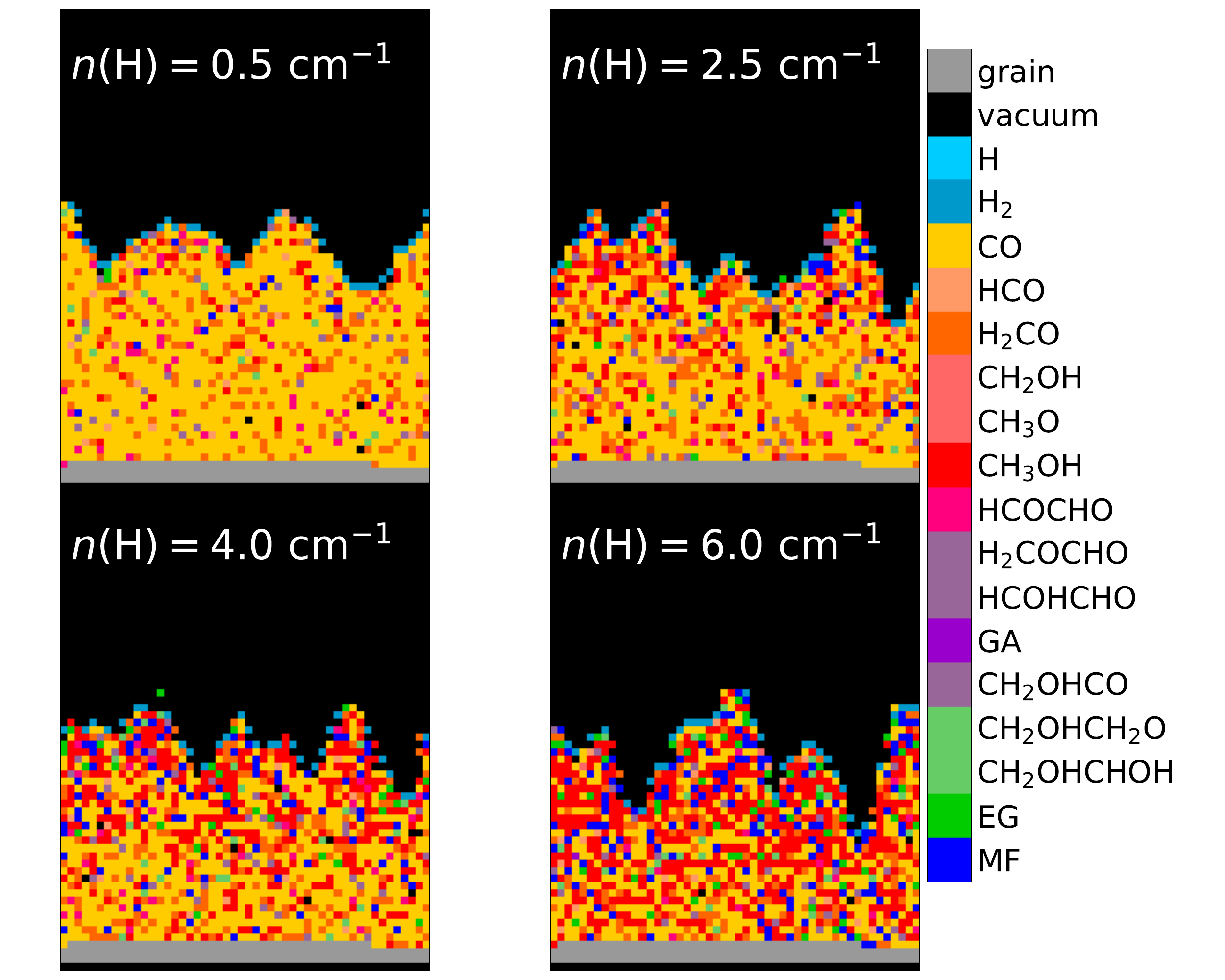}
        \caption{Cross sections of the formed ice layer for four different levels of hydrogen abundances: 1.0, 2.5, 4.0, and 6.0~cm$^{-3}$. The abundance of gas-phase carbon monoxide is 10.0~cm$^{-3}$. The temperature of the grain is 10~K. }
        \label{fig:ISM_grain}
\end{figure}

Figure~\ref{fig:ISM_time} shows the evolution of a grain surface for four different hydrogen abundances in dark molecular cloud conditions. These simulated abundances are $n(\text{H}) =$ 1.0, 2.5, 4.0 and 6.0~cm$^{-3}$ in panels a through d, respectively, which span an extreme range in accordance with \citet{Goldsmith:2005}, who found $n(\ce{H})\approx2$--$6$ cm$^{-3}$. The initial CO abundance is $n_\text{initial}(\text{CO}) =$ 10~cm$^{-3}$. This relates to a density of $n_\text{H}= 10^{5}$~cm$^{-3}$ using the typical relative abundance of 10$^{-4}$ \citep{Bergin:2007}, and the grain-surface temperature is equal to 10~K in all cases. The left-hand vertical axis (red) denotes the grain-surface build-up in number of particles per surface area, whereas the right-hand axis (blue) denotes the grain-surface abundance with respect to $n_\text{H}$. The total simulated time is $2\cdot10^5$ years for all ISM simulations (Table~\ref{tab:simulations}).

Carbon monoxide freeze-out is faster than hydrogenation at the initial stage, as shown in Fig.~\ref{fig:ISM_time}. Some of this carbon monoxide is hydrogenated into formyl radicals (HCO) and subsequently into formaldehyde (\ce{H2CO}). \ce{CO} remains dominant for the simulated time in all four simulations. A small decrease in mantle build-up can be seen toward the end of the simulations; this is a result of CO depletion from the gas phase. In all simulations hydrogenation and covering of the surface by incoming \ce{CO} molecules compete.

Panel a of Fig.~\ref{fig:ISM_time} is the result of a simulation with a low H:CO ratio where the covering by \ce{CO} is fast with respect to hydrogenation. The mantle is mainly composed of CO, with \ce{H2CO} being the second most abundant species. A cross section of the formed grain mantle can be found in Fig.~\ref{fig:ISM_grain}, which shows the formation of compact CO ice with  some minority species, predominantly \ce{H2CO}, and covered with a layer of \ce{H2}. In these conditions the formation of \ce{HCO} is relatively rare, and two \ce{HCO} radicals are hardly formed in close proximity. The low amount of glycolaldehyde that forms indicates that few HCO recombinations have occurred. Enough hydrogen is present to further efficiently hydrogenate \ce{HCO} to \ce{H2CO} . In the latter stages of the simulation, the efficiency of glycolaldehyde formation is sufficient to form ethylene glycol because the CO deposition rate decreases as a result of progressed depletion in the gas phase.

Results of simulations with a slightly higher gas-phase hydrogen are shown in Fig.~\ref{fig:ISM_time}b and in the upper right panel of Fig.~\ref{fig:ISM_grain}. The grain mantle now features less CO in favor of \ce{H2CO} and more complex species. COMs are present on the grain early in the simulation. Because the H:CO ratio is higher than in the previously described simulation, the formation of HCO has increased and alongside the probability that two formyl radicals are formed in close proximity. The formation of glycolaldehyde and ethylene glycol, which are the result of subsequent hydrogenation, competes with direct hydrogenation of HCO toward methanol. These effects increase when the abundance of gas-phase hydrogen is raised to $n(\text{H}) = 4.0 \text{~cm}^{-3}$ (Fig.~\ref{fig:ISM_time}c).

At the highest simulated hydrogen abundance $n(\text{H}) = 6.0 \text{~cm}^{-3}$ , the amount of methanol equals the amount of formaldehyde, and EG is more abundant than GA. Moreover, methyl formate formation generally increases with hydrogen abundance. This is to be expected because it is dependent on both HCO and \ce{CH3O}, whose formation also increases with the presence of hydrogen.

\begin{figure}[tb!]
        \includegraphics[width=0.45\textwidth]{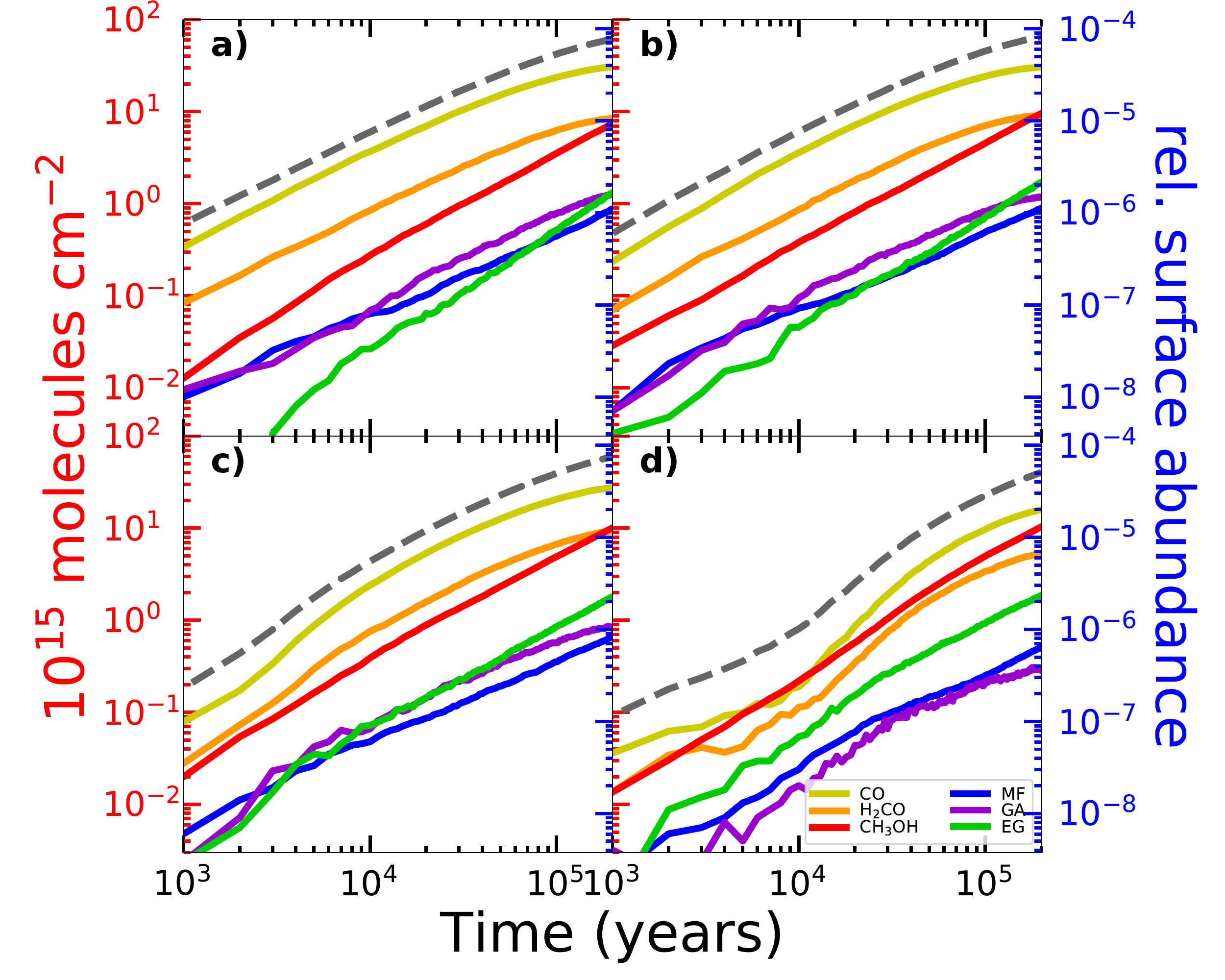}
        \caption{Simulated time evolution of the grain surface composition for four different temperatures: 8, 16, 18, and 20~K (a-d resp.). The abundance of gas-phase carbon monoxide is 10.0~cm$^{-3}$ and of hydrogen atoms is 2.5~cm$^{-3}$. }
        \label{fig:ISM_temp}
\end{figure}

\begin{figure}[tb!]
        \includegraphics[width=0.45\textwidth]{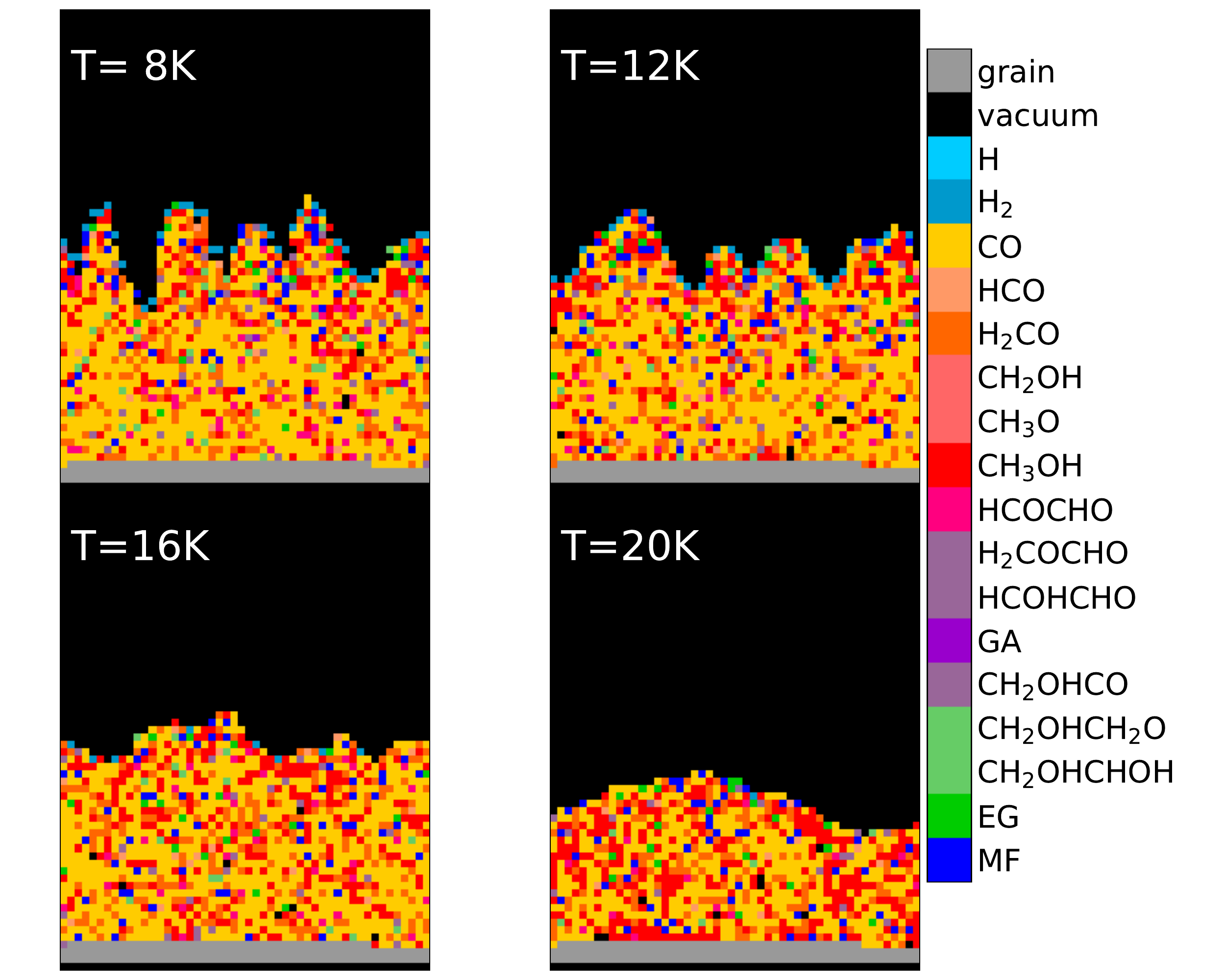}
        \caption{Simulated mantle cross sections for four different temperatures: 8, 12, 16, and 20~K. The abundance of gas-phase carbon monoxide is 10.0~cm$^{-3}$ and of hydrogen atoms is 2.5~cm$^{-3}$. }
        \label{fig:ISM_temp_grain}
\end{figure}

To study the effect of temperature, the simulations were repeated for 8, 12, 14, 16, 18, and 20~K. Figure~\ref{fig:ISM_temp} shows the mantle composition for a selection of these simulations: $n(\text{H}) =$ 2.5~cm$^{-3}$ and $n_\text{initial}(\text{CO}) =$ 10~cm$^{-3}$ and four different temperatures: a) 8~K, b) 12~K, c) 16~K, and d) 20~K. The surface abundance of the main species is rather independent of the surface temperature in the range 8--16~K. Because we took the gas and grain temperature to be equal, the deposition rate increased with temperature ($\propto \sqrt{T}$). This has some effect on the H:CO ratio at late times because CO depletes faster at 16~K than at 8~K.  \ce{H2CO}/\ce{CH3OH} and \ce{GA}/\ce{EG} decrease with temperature as a consequence. 
Figure~\ref{fig:ISM_temp_grain} shows the mantle cross sections corresponding to the results in Fig.~\ref{fig:ISM_temp}. The build-up of ice at 20~K is initially significantly reduced due to desorption of CO at these elevated temperatures, but the formation of \ce{CH3OH} helps to retain some CO at later times. Comparison between 8, 10 (Figs.~\ref{fig:ISM_temp}), 12, and 16~K shows that the surface structure of the ice is smoother for higher temperatures because of the increased mobility of CO. Where the surface is covered with \ce{H2} for 8 to 12~K, this has all disappeared at 16~K. A full overview of all the obtained results of dark cloud simulations can be found in Table \ref{tab:ISM_results}.

\renewcommand{\baselinestretch}{1}
\begin{sidewaystable*}
\centering
\caption{Grain surface abundance of CO-hydrogenation products at the end of simulations. Results are shown with respect to methanol.}
\label{tab:ISM_results}
  \begin{tabular}{l l l r r r r r r r r r r}
        \hline\hline    
                & $T$ [K]       & $n(\text{H})$ [cm$^{-3}$]     & CO    & HCO     & \ce{H2CO}     & \ce{H2COH}    & \ce{H3CO}     & GX    & GA            & EG              & MF            & monolayers    \\
        \hline
        7A   &  {   }8    & 1.0   & 19.219    & 0.390 & 3.492 & 0.002 & 0.008 & 0.921 & 0.334 & 0.099 & 0.144 & 29.587\\
        7B   &  {   }8    & 2.5   & 4.157     & 0.092 & 1.148 & 0.002 & 0.003 & 0.224 & 0.171 & 0.179 & 0.119 & 28.016\\
        7C   &  {   }8    & 4.0   & 1.974     & 0.053 & 0.599 & 0.002 & 0.003 & 0.109 & 0.102 & 0.209 & 0.112 & 26.946\\
        7D   &  {   }8    & 6.0   & 1.087     & 0.036 & 0.340 & 0.002 & 0.005 & 0.059 & 0.065 & 0.218 & 0.116 & 26.032\\
        8A   & 10       & 1.0   & 16.064    & 0.332 & 3.013 & 0.002 & 0.007 & 0.774 & 0.290 & 0.101 & 0.119 & 30.301\\
        8B   & 10       & 2.5   & 3.574     & 0.083 & 1.007 & 0.003 & 0.002 & 0.210 & 0.155 & 0.175 & 0.113 & 28.463\\
        8C   & 10       & 4.0   & 1.706     & 0.044 & 0.532 & 0.002 & 0.003 & 0.100 & 0.097 & 0.206 & 0.103 & 27.462\\
        8D   & 10       & 6.0   & 0.921     & 0.032 & 0.302 & 0.003 & 0.004 & 0.059 & 0.062 & 0.222 & 0.109 & 26.400\\
        9A   & 12       & 1.0   & 14.959    & 0.292 & 3.020 & 0.003 & 0.004 & 0.676 & 0.269 & 0.131 & 0.118 & 30.537\\
        9B   & 12       & 2.5   & 3.181     & 0.077 & 0.945 & 0.002 & 0.002 & 0.171 & 0.124 & 0.177 & 0.093 & 29.255\\
        9C   & 12       & 4.0   & 1.536     & 0.042 & 0.512 & 0.002 & 0.003 & 0.086 & 0.079 & 0.198 & 0.088 & 28.186\\
        9D   & 12       & 6.0   & 0.820     & 0.031 & 0.291 & 0.001 & 0.003 & 0.049 & 0.050 & 0.207 & 0.094 & 27.240\\
        10A  & 14       & 1.0   & 13.582    & 0.250 & 2.829 & 0.002 & 0.004 & 0.570 & 0.292 & 0.138 & 0.093 & 30.422\\
        10B  & 14       & 2.5   & 3.050     & 0.072 & 0.955 & 0.001 & 0.002 & 0.158 & 0.121 & 0.191 & 0.082 & 29.378\\
        10C  & 14       & 4.0   & 1.470     & 0.044 & 0.508 & 0.001 & 0.003 & 0.081 & 0.075 & 0.198 & 0.082 & 28.310\\
        10D  & 14       & 6.0   & 0.776     & 0.029 & 0.291 & 0.002 & 0.003 & 0.046 & 0.046 & 0.205 & 0.086 & 27.384\\
        11A  & 16       & 1.0   & 11.057    & 0.188 & 2.612 & 0.002 & 0.002 & 0.340 & 0.209 & 0.149 & 0.074 & 26.721\\
        11B  & 16       & 2.5   & 2.748     & 0.064 & 0.933 & 0.001 & 0.002 & 0.107 & 0.083 & 0.178 & 0.066 & 27.024\\
        11C  & 16       & 4.0   & 1.326     & 0.039 & 0.497 & 0.001 & 0.001 & 0.060 & 0.050 & 0.175 & 0.066 & 26.701\\
        11D  & 16       & 6.0   & 0.712     & 0.028 & 0.283 & 0.001 & 0.002 & 0.036 & 0.031 & 0.178 & 0.065 & 26.390\\
        12A  & 18       & 1.0   & 7.352     & 0.126 & 1.961 & 0.000 & 0.000 & 0.174 & 0.088 & 0.157 & 0.052 & 19.248\\
        12B  & 18       & 2.5   & 2.668     & 0.054 & 0.879 & 0.001 & 0.002 & 0.087 & 0.053 & 0.177 & 0.056 & 24.152\\
        12C  & 18       & 4.0   & 1.315     & 0.032 & 0.464 & 0.001 & 0.002 & 0.052 & 0.035 & 0.168 & 0.058 & 25.226\\
        12D  & 18       & 6.0   & 0.717     & 0.025 & 0.268 & 0.000 & 0.003 & 0.031 & 0.021 & 0.172 & 0.061 & 25.311\\
        13A  & 20       & 1.0   & 4.387     & 0.075 & 1.297 & 0.001 & 0.001 & 0.115 & 0.063 & 0.197 & 0.047 & 13.758\\
        13B  & 20       & 2.5   & 1.542     & 0.035 & 0.520 & 0.000 & 0.002 & 0.061 & 0.028 & 0.180 & 0.049 & 18.111\\
        13C  & 20       & 4.0   & 0.827     & 0.023 & 0.284 & 0.001 & 0.002 & 0.034 & 0.020 & 0.173 & 0.054 & 19.220\\
        13D  & 20       & 6.0   & 0.429     & 0.016 & 0.147 & 0.001 & 0.003 & 0.022 & 0.014 & 0.170 & 0.060 & 19.521\\
        \hline\hline
  \end{tabular}
\end{sidewaystable*}
\renewcommand{\baselinestretch}{1.5}

\subsection{Formation routes}
\begin{figure*}
 \includegraphics[width=\textwidth]{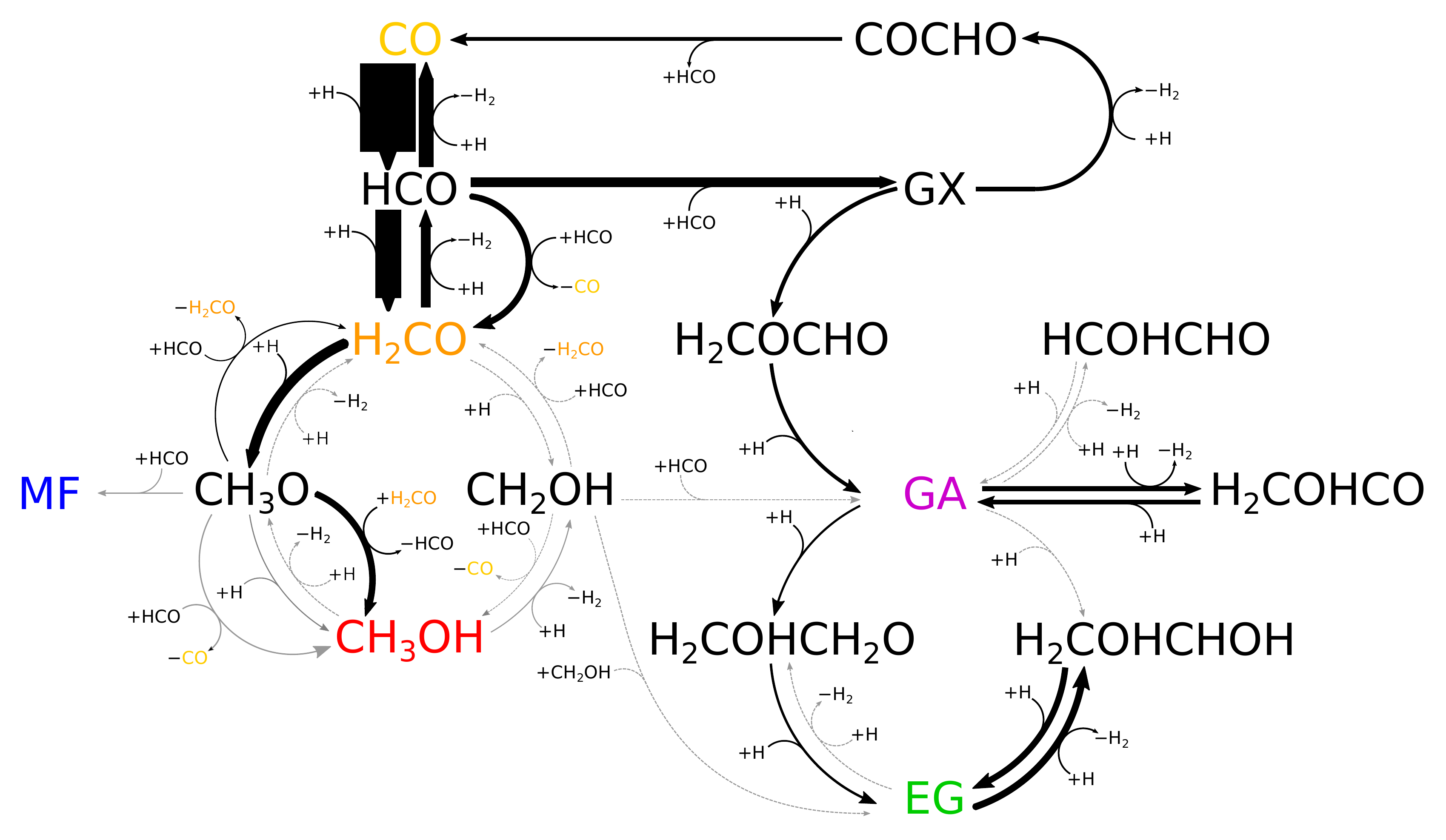}
 \caption{Flux distribution of the network for a standard simulation of $n(\ce{CO}) = 10.0$~cm$^{-3}$, $n(\ce{H}) =2.5$~cm$^{-3}$ and $T=10$~K. The thickness of the arrows indicates the relative frequency of the reaction. Some arrows become too thin and are indicated in gray. Reactions that contribute less than 0.1 \% with respect to \ce{H + CO -> HCO} are dashed. }
 \label{fig:flux}
\end{figure*}

Figure~\ref{fig:flux} shows  the flux distribution of the network for a standard simulation of $n(\ce{CO}) = 10.0$~cm$^{-3}$, $n(\ce{H}) =2.5$~cm$^{-3}$ , and $T=10$~K. The thickness of the arrows indicates the relative frequency of the reaction. The figure clearly shows that HCO constantly reacts and reforms. In this process of hydrogenation and abstraction, \ce{H2} is formed, which becomes the dominant route to form \ce{H2}.  This holds for the ISM and experimental simulations. Hydrogen atoms react away before a new hydrogen atom can land on the surface, and the lifetime of a single H atom on the surface is too short to form \ce{H2} through \ce{H + H -> H2}. Hydrogenation reactions are hence the main loss route for H atoms. This also explains the weak dependence of the results in Fig.~\ref{fig:all_exp} on the binding energy of H to the surface. 

Furthermore, the recycling of EG is immediately obvious. The rate constant for the abstraction reaction of EG is high, much higher than for GA and \ce{CH3OH}. The \ce{H2COHCHOH} radical that is formed in this way has its unpaired electron on a carbon. Reaction of this species with HCO could further elongate toward glycerol. 

The formation routes of methanol and glycoaldehyde both require at least four hydrogenation steps. Methanol is the more abundantly formed species of the two for all conditions. The majority of methanol is formed from the reaction between \ce{CH3O} and \ce{H2CO}, as we show in Fig.~\ref{fig:flux}. This is further quantified in Table~\ref{tab:formroutes} for the ISM and experimental conditions. The relative contributions of the different formation routes of \ce{H2CO} and \ce{CH3OH} was found to be relatively insensitive to the temperature, but dependent on the H:CO ratio. The table therefore only shows the contribution per simulated H:CO ratio for 10~K. 
In all cases, \ce{CH3O + H2CO -> CH3OH + HCO} is the dominant formation route for methanol and not \ce{CH3O + H -> CH3OH} , which might be intuitively expected to be the dominant formation route. \citet{Penteado:2017} have shown for a full gas-grain model with diffusive surface reactions that methanol is predominantly formed from destruction reactions of more complex species rather than through hydrogenation of \ce{CH3O}. The reaction \ce{CH3O + H2CO -> CH3OH + HCO} was underestimated in this study because radical-radical recombination without diffusion was not taken into account. In both cases, the reason is the low surface coverage by H atoms. In the present case, H atoms mostly react with CO, HCO, \ce{H2CO} , and EG and its radical to produce \ce{H2}. If two formaldehyde molecules are formed close to each other,  only one hydrogen atom is needed for hydrogenation to \ce{CH3O} to convert this into \ce{CH3OH} and HCO. The latter will be quickly hydrogenated again. \ce{CH3O} radicals that are not formed in close proximity to a \ce{H2CO} molecule will be quickly hydrogenated by one of the \ce{H2} molecules that covers the surface. Only at higher temperatures, when the surface coverage of \ce{H2} is low, does the reaction \ce{CH3O + H -> CH3OH} start to play a role in methanol formation.
The other product branch \ce{CH3O + H -> H2CO + H2} leading to formaldehyde is similarly only important for elevated temperatures. \ce{H2CO} is predominantly formed through hydrogenation of HCO, with \ce{HCO + HCO -> H2CO + CO} of increasing importance with lower H:CO ratio.

\begin{table*}
\centering
\caption{Contribution of the individual reactions to the formation of formaldehyde and methanol}
\label{tab:formroutes}
\begin{tabular}{l@{ + }l|rrr|rrrr}
\hline
\hline
\multicolumn{2}{l}{}       & \multicolumn{3}{l|}{\textbf{Experimental conditions}}                          & \multicolumn{4}{l}{\textbf{Interstellar conditions} (10~K)}    \\
\multicolumn{2}{l}{}       & \multicolumn{2}{c}{\ce{H + CO}} & \ce{H + H2CO} & \multicolumn{4}{c}{\ce{H + CO}} \\    
\multicolumn{2}{l}{}       & 14 K                 & 15 K                   & 15 K                           & 1.0 cm$^{-3}$& 2.5 cm$^{-3}$& 4.0 cm$^{-3}$& 6.0 cm$^{-3}$\\
\hline
\multicolumn{2}{l}{\ce{H2CO} formation}                                                             \\ 
\hline
\ce{HCO}   & \ce{H}      & 73.86\% & 73.19\% & 60.78\%          & 70.13\% & 73.63\% & 75.22\% & 76.48\% \\
\ce{CH3O}  & \ce{H}      & 1.37\%  & 1.31\%  & 0.68\%           & 0.58\%  & 0.94\%  & 1.28\%  & 1.90\%  \\
\ce{HCO}   & \ce{HCO}    & 17.42\% & 19.54\% & 16.88\%          & 22.99\% & 17.59\% & 14.59\% & 12.28\% \\
\ce{CH2OH} & \ce{HCO}    & 0.08\%  & 0.02\%  & 0.11\%           & 0.04\%  & 0.13\%  & 0.28\%  & 0.34\%  \\
\ce{CH3O}  & \ce{HCO}    & 7.04\%  & 5.78\%  & 20.97\%          & 6.20\%  & 7.48\%  & 8.12\%  & 8.18\%  \\
\ce{CH2OH} & \ce{CH3O}   & 0.02\%  & 0.02\%  & 0.26\%           & 0.01\%  & 0.11\%  & 0.28\%  & 0.49\%  \\
\ce{CH3O}  & \ce{CH3O}   & 0.21\%  & 0.14\%  & 0.32\%           & 0.04\%  & 0.12\%  & 0.24\%  & 0.33\%  \\
\hline
\multicolumn{2}{l}{\ce{CH3OH} formation}                                                            \\
\hline
\ce{CH2OH} & \ce{H}      & 2.72\%  & 1.77\%  & 1.13\%           & 1.42\%  & 3.91\%  & 7.36\%  & 12.32\% \\
\ce{CH3O}  & \ce{H}      & 12.62\% & 13.66\% & 1.49\%           & 7.44\%  & 6.62\%  & 8.62\%  & 11.08\% \\
\ce{CH2OH} & \ce{HCO}    & 1.24\%  & 0.54\%  & 0.50\%           & 1.25\%  & 2.72\%  & 3.63\%  & 4.84\%  \\
\ce{CH3O}  & \ce{HCO}    & 9.51\%  & 9.80\%  & 6.06\%           & 10.00\% & 8.30\%  & 6.81\%  & 6.48\%  \\
\ce{CH3O}  & \ce{H2CO}   & 72.70\% & 73.33\% & 90.14\%          & 79.58\% & 77.50\% & 71.82\% & 62.85\% \\
\ce{CH2OH} & \ce{CH3O}   & 0.12\%  & 0.11\%  & 0.31\%           & 0.09\%  & 0.45\%  & 0.95\%  & 1.45\%  \\
\ce{CH3O}  & \ce{CH3O}   & 1.09\%  & 0.80\%  & 0.37\%           & 0.23\%  & 0.51\%  & 0.81\%  & 0.97\%  \\
\hline 
\hline
\end{tabular}
\end{table*}

\subsection{Comparison with observations}
To place our simulations into astrophysical context, we would like to compare our results to astrophysical observations. Unfortunately, this cannot be done directly to ice observations. Complex species such as EG, GA, and MF are hard to detect with IR observations because of sensitivity and selectivity issues. They have been observed in the gas phase toward various astrophysical objects. Table~\ref{tab:ISM_observations} shows a selection of observed abundances of \ce{H2CO}, GA, EG, and MF with respect to \ce{CH3OH}. For cometary observations, this corresponds to the coma observations where mantle material desorbs by sunlight. For hot corinos and hot cores, the temperature of the grains in the object is sufficiently high to have lost all its mantle material. Finally, for the cold and quiescent sources, the ice mantles are mostly intact and the observed gas-phase species have desorbed from the grain through some nonthermal desorption mechanism such as photodesorption, reactive desorption, and/or cosmic-ray desorption (either through heating or sputtering). In all cases, we assume that the gas-phase composition reflects the composition of the original ice mantle. This only holds if gas-phase chemistry has not sufficiently altered their abundance. This is especially relevant for \ce{H2CO} , which also has gas-phase formation routes. In cometary coma, gas-phase chemistry is unlikely to occur. Because for cold cloud observations not all ice has desorbed, we need to take into account that the nonthermal desorption efficiencies are not the same for all species. The observed composition is therefore not a direct reflection of the mantle composition. The observations will give us some guidance on the accuracy of our model, however, and might indicate the physical conditions under which the ice mantle has been formed.

The \ce{H2CO}/\ce{CH3OH} ratio in Table~\ref{tab:ISM_observations} shows large variation, but remains below unity for all cases. Comparison with Table~\ref{tab:ISM_results} shows that this means that the H:CO ratio should be at least 0.3. Quiescent clouds toward the Galactic center show an even lower \ce{H2CO}/\ce{CH3OH}. This is hard to reconcile directly with our results, but might be caused by a higher reactive desorption efficiency for \ce{CH3OH} than for \ce{H2CO}. Methanol is predominantly formed in a reaction with two products, whereas an addition reaction is the most dominant route for \ce{H2CO} formation. Bond-breaking reactions are more likely to lead to translational excitation than to an addition reaction. Translational excitation can result in very efficient desorption \citep{Fredon:2017, Fredon:2018}. Missing \ce{CH3OH} formation routes could be another reason, for instance, \ce{CH3O + HCOOH -> CH3OH + COOH}.

The GA/EG ratio also varies strongly. In our simulations, \ce{H2CO}/\ce{CH3OH} and GA/EG ratios are directly correlated because both \ce{CH3OH} and EG formation at the expense of \ce{H2CO} and GA, respectively, requires a high H abundance. In the observational data, no clear correlation between these two ratios can be observed. The reason may be the limited data set with an inhomogeneity of sources. The simulated EG and GA abundance is higher than the observed abundance. This could be due to missing EG destruction routes, as discussed in the previous section, or to missing \ce{CH3OH} formation routes through \ce{CH3 + OH}. 

Methylformate has abundances of a few percent with respect to methanol for all sources except for the hot core source, where MF is more abundant than \ce{CH3OH}. The low abundances are consistent with our simulations, where a rather constant MF abundance with respect to \ce{CH3OH} can be observed. The high abundance of MF observed in the hot core is probably due to a different formation mechanisms. Our simulations are representative for a cold and dark grain chemistry, whereas during high-mass star formation, the cloud is probably warmer and more exposed to UV light.
This leads to more radicals and a higher mobility of these radicals. This favors radical-radical recombination to form MF, which is more in line with the simulations by \citet{Garrod:2006a}. The MF/\ce{CH3OH} might serve as a probe for cold dark grain chemistry.

\begin{table*}
\centering
\caption{Summary of observational data of various dark interstellar environments. Key species are shown with respect to methanol where available.}\label{tab:ISM_observations}
  \begin{minipage}{0.55\textwidth}
  \begin{tabular}{l c c c c l}
   \hline\hline
    Source                              & \ce{H2CO}     & GA                            & EG                              & MF                            & ref                                                                   \\
   \hline
    \multicolumn{6}{c}{comet} \\
    67p                                 & 0.9636        & $-$                           & \textless 0.0045        & 0.0418                        & 1                                             \\
    Hale-Bopp                   & 0.4583        & \textless 0.0166      & 0.1042                  & 0.0333                        & 2,3   \\
    Lemmon                              & 0.4375        & \textless 0.05        & 0.15                            & \textless 0.1         & 4                                             \\
    Lovejoy                             & 0.2692        & \textless 0.0269      & 0.1346                  & \textless 0.0769      & 4                                             \\
   \hline
    \multicolumn{6}{c}{low-mass star-forming region (hot corino)} \\
    IRAS~16293-2422             & 0.0821        & 0.0027                        & 0.01                            & 0.02                          & 5,6,7 \\
    NGC~1333-IRAS~2A    & 0.5880        & 0.0065                        & 0.0023                  & 0.019                         & 5,8,9 \\    NGC~1333-IRAS~4A  & 0.2990  & 0.0141                        & 0.0025                        & 0.015                           & 5,9\\
   \hline
    \multicolumn{6}{c}{high-mass star-forming region (hot core)} \\
    G31.41+0.31                 & $-$           & 0.1333                        & 1.333                           & 4.5333                        & 10,11         \\
   \hline
    \multicolumn{6}{c}{dark prestellar core} \\
    Barnard 1b                  & $-$           & \textless 0.011       & $-$                             & 0.012                         & 12                                            \\
   \hline
    \multicolumn{6}{c}{Galactic center quiescent cloud} \\
    G$-$0.02                    & 0.053         & 0.010                         & 0.012                           & 0.034                         & 13            \\
    G$-$0.11                    & 0.059         & 0.015                         & 0.020                           & 0.071                         & 13            \\
    G+0.693                             & 0.020         & 0.020                         & 0.024                           & 0.10                          & 13            \\
   \hline\hline
  \end{tabular}  
  
  \begin{tabular}{l l}
        $^1$ \citet{Leroy:2015}                         & $^8$ \citet{Coutens:2015}\\
        $^2$ \citet{Bockelee-Morvan:2000}       & $^9$ \citet{Taquet:2015}\\
        $^3$ \citet{Crovisier:2004}                     & $^{10}$ \citet{Beltran:2008}\\
        $^4$ \citet{Biver:2014}                         & $^{11}$ \citet{Rivilla:2017}\\
        $^5$ \citet{Bottinelli:2007}            & $^{12}$ \citet{Oberg:2010}\\
        $^6$ \citet{Jorgensen:2016}                     & $^{13}$ \citet{Requena:2008}\\
        $^7$ \citet{Rivilla:2018} &\\
  \end{tabular}

  \end{minipage}

\end{table*}

\section{Conclusions}
In summary, the present paper presented Monte Carlo simulations to constrain methyl formate, glycolaldehyde, and ethylene glycol formation on the surface of interstellar dust grains in  cold and dark molecular clouds. For this purpose, a new CO-hydrogenation reaction network was presented. Using this network, we arrived at the following conclusions.

\begin{enumerate}
\item{The rate constants in this network are based on literature values. The constants were calculated using harmonic quantum transition state theory. This takes hydrogen tunneling correctly into account. The branching ratios for radical-radical reactions in this new network were determined quantum-chemically as well. We present a large consistent set of these values.}
\item {This new network was benchmarked against experimental results of \cite{Chuang:2016, Chuang:2017} on CO and \ce{H2CO} hydrogenation. A good agreement is reached with this new model.}
\item{A comparison between experiments and simulations using rate constants and branching ratios from UMIST showed an overproduction of \ce{CH3OH} , which stresses the importance of considering both hydrogenation and abstraction reactions instead of effective hydrogenation rates.}
\item{The new model was used to simulate CO freeze-out in dense molecular clouds. A grid of simulations was performed with varying dust temperatures and H-to-CO ratios. The final COM abundances depend more on the H/CO and less on temperature. Only above 14~K, where CO build-up is less efficient, does temperature start to play a role.}
\item{Production of GA and EG through HCO recombination occurs at all H/CO and temperatures we studied, even at temperatures as low as 8~K. The reason is that the \ce{HCO + HCO} reaction can occur when HCO radicals are formed close to each other and does not require diffusion of the radicals. Relatively low abundances of MF have been formed. }
\item{Molecular hydrogen is predominantly formed through abstraction reactions on the surface. The most  important reaction leading to methanol is \ce{H2CO + CH3O -> HCO + CH3OH}.}
\item{Our simulations agree reasonably well with observed COM ratios for mantles that have been formed at low temperatures. }
\end{enumerate}

\section{Acknowledgements}
M.S. is grateful for financial support from the Dutch Astrochemistry Network II (DANII) financed by the Dutch Organisation for Scientific Research (NWO). T.L. is grateful for support from NWO via a VENI fellowship (722.017.008). We would like to thank Naoki Watanabe for his contribution to fruitful discussions.

\end{document}